\documentclass[journal]{IEEEtran}

\usepackage{epsfig,amsmath,amssymb,epsf,amsthm,scalefnt,multirow}
\usepackage{xcolor}
\usepackage{float}
\usepackage{cite}
\usepackage{psfrag}
\usepackage{gensymb}
\usepackage{cleveref}
\usepackage{mathrsfs}
\usepackage{mathtools}
\usepackage{algpseudocode}
\usepackage{algorithm}
\usepackage{enumerate}
\usepackage{stfloats}
\usepackage{latexsym}
\usepackage{multicol}
\usepackage[caption=false]{subfig}

\newtheorem*{lemma*}{Lemma}

  \def\cC{{\mathcal{C}}} 
   
  \def\cK{{\mathcal{K}}} 
\def\cM{{\mathcal{M}}} \def\cN{{\mathcal{N}}} \def\cO{{\mathcal{O}}} 
 \def\cR{{\mathcal{R}}}  \def\cT{{\mathcal{T}}}

\def\diag{\mathop{\mathrm{diag}}}

\def\bPhi{{\pmb{\Phi}}} \def\bphi{{\pmb{\phi}}}

\def\b0{{\pmb{0}}} 

   \def\bd{{\mathbf{d}}}
 \def\bff{{\mathbf{f}}}  \def\bh{{\mathbf{h}}}

  \def\bs{{\mathbf{s}}}

\def\bA{{\mathbf{A}}}   
 \def\bF{{\mathbf{F}}} \def\bG{{\mathbf{G}}} 
\def\bI{{\mathbf{I}}}   
   
 \def\bR{{\mathbf{R}}}

\begin{document}

\title{RIS-Enabled Cellular Systems \\ Operated by Different Service Providers}

\author{Hyeongtaek~Lee, \textit{Member,~IEEE},
	and~Junil~Choi, \textit{Senior Member, IEEE} 

 \thanks{Hyeongtaek~Lee and Junil~Choi are with the School of Electrical Engineering, Korea Advanced Institute of Science and Technology (e-mail: \{htlee8459; junil\}@kaist.ac.kr).}
}

\maketitle

\begin{abstract}
In realistic cellular communication systems, multiple service providers will operate within different frequency ranges. Each serving cell, which is managed by a distinct service provider, is designed individually due to the orthogonal frequencies. However, when a reconfigurable intelligent surface (RIS) is deployed for a certain cell, the RIS still incurs reflective channels for the overall system since the RIS reflects signals across all frequency ranges. This may cause severe undesired performance degradation for the other cells unless the reflection coefficients are properly designed. To tackle this issue, by utilizing the Riemannian manifold optimization method, an RIS reflection coefficients design is proposed in this paper to maximize the performance improvements of the cell that deploys the RIS while minimizing the undesired performance degradation for the other cells simultaneously. Numerical results demonstrate that the proposed design can effectively balance the two objectives for practical scenarios. 
\end{abstract}

\section{Introduction} \label{section: introduction}

As a key enabling technology to realize a smart radio environment, utilizing reconfigurable intelligent surfaces (RISs) is drawing great attention in recent years \cite{RIS_intro_1, RIS_intro_2, RIS_intro_3}. With a massive number of passive elements, which can induce favorable phase shifts to incoming signals, the RIS can construct controllable wireless channels and improve the communication system in various ways, e.g., increased spectral efficiency and reduced power consumption. The RIS also can be utilized to assist the communications with unmanned aerial vehicles or high-altitude platforms \cite{UAV_HAP_RIS_1, UAV_HAP_RIS_2, UAV_HAP_RIS_3}, satellite-terrestrial relay systems \cite{NTN_RIS}, and internet-of-things networks \cite{IoT_RIS}.

In addition to the above applications, some existing works considered the RIS-assisted multi-cell communication systems \cite{multi_cell_RIS_1,multi_cell_RIS_2,multi_cell_RIS_3,multi_cell_RIS_4,mutli_cell_RIS_5,multi_cell_RIS_6_selective}. In \cite{multi_cell_RIS_1}, the RIS was deployed to maximize the weighted sum-rate of all users by assisting the cell-edge users and mitigating the inter-cell interference. Another sum-rate maximization problem was considered in \cite{multi_cell_RIS_2} where the non-orthogonal-multiple-access (NOMA) users located at the cell-edges are supported by the RIS. The RIS was developed to maximize the geometric mean of signal-to-interference-plus-noise-ratio (SINR) when the users are served by two base stations (BSs) in \cite{multi_cell_RIS_3}. To suppress the inter-cell interference, the reflection coefficients of RIS were designed to maximize the minimum weighted SINR in \cite{multi_cell_RIS_4}. By utilizing multiple RISs, the total time-frequency resource consumption minimization problem under demand requirement was formulated in \cite{mutli_cell_RIS_5}. Although these works made noticeable performance improvements, they were limited to the single service provider case where the users in multiple cells are served by the same service provider through the same operating frequency range. While different frequency ranges were assumed for multiple cells in \cite{multi_cell_RIS_6_selective}, the work was also restricted to the case when the multiple BSs belong to the same service provider where the RIS can be designed to improve the performance of overall system, e.g., the sum-rate of users over multiple cells. Note that deploying the RIS to enhance the overall communication system performance is possible only when the multiple cells are operated by the same service provider and when the BSs can cooperate. However, in scenarios where each cell is operated by a distinctive service provider, the most practical strategy for utilizing an RIS would be to minimize or neutralize its impact on neighboring cells operated by different service providers.

In this paper, we consider a realistic cellular communication system where each BS, which is operated by a certain service provider, supports the users in its serving cell using a separate frequency range. Because of the orthogonal frequencies, any interaction, such as inter-cell interference, among different cells does not exist at all. When a service provider deploys an RIS for its serving cell, however, the RIS affects the channels of the other cells since RISs reflect signals across all frequency ranges. Therefore, while there is no inter-cell interference, users in the other cells inevitably receive signals from not just their serving BSs but also from the RIS that reflects any incoming signals. This can cause undesired and severe performance degradation for the overall system unless the RIS is designed considering its effect on the other~cells. Although not common in literature, investigating the effect of RIS for this practical scenario will be important, and there has been no related prior work to the best of our knowledge. To tackle this scenario, we carefully design the reflection coefficients of RIS by exploiting the Riemannian manifold optimization method to balance between the two objectives: 1) maximizing the performance improvements of the cell that deploys the RIS, and 2) minimizing the performance degradation of the other cells. Numerical results clearly show the effectiveness of proposed design for practical scenarios. 

The paper is organized as follows. We explain the system model of scenario of interest in Section \ref{section: system model}. Then, in Section \ref{section: proposed balancing reflection design}, the proposed balancing RIS design is developed. Numerical results are shown in Section \ref{section: numerical results} to evaluate the performance of the proposed design, and we conclude the paper in Section~\ref{section: conclusion}.

\textit{Notations}: Lower and upper boldface letters denote column vectors and matrices. The transpose, conjugate transpose, and element-wise conjugate of a matrix $\bA$ are represented by $\bA^\mathrm{T}$, $\bA^\mathrm{H}$, and $\bA^*$. For a square matrix $\bA$, $\mathrm{Tr}(\bA)$ is the trace of $\bA$. The diagonalization operation is denoted by $\diag(\cdot)$. Notation $\cC\cN(\mu, \sigma^2)$ stands for the complex Gaussian distribution with mean $\mu$ and variance $\sigma^2$. For a matrix $\bA$, $\Vert \bA \Vert_\mathrm{F}$ is the Frobenius norm of $\bA$. Notation $\mathrm{Re}(a)$ represents the real part of a complex number $a$, the Hadamard product is denoted by~$\odot$, and $\bI_m$ is the $m \times m$ identity matrix. 

\section{System Model} \label{section: system model}

\begin{figure}[t]
	\centering
	\includegraphics[width=0.9\columnwidth]{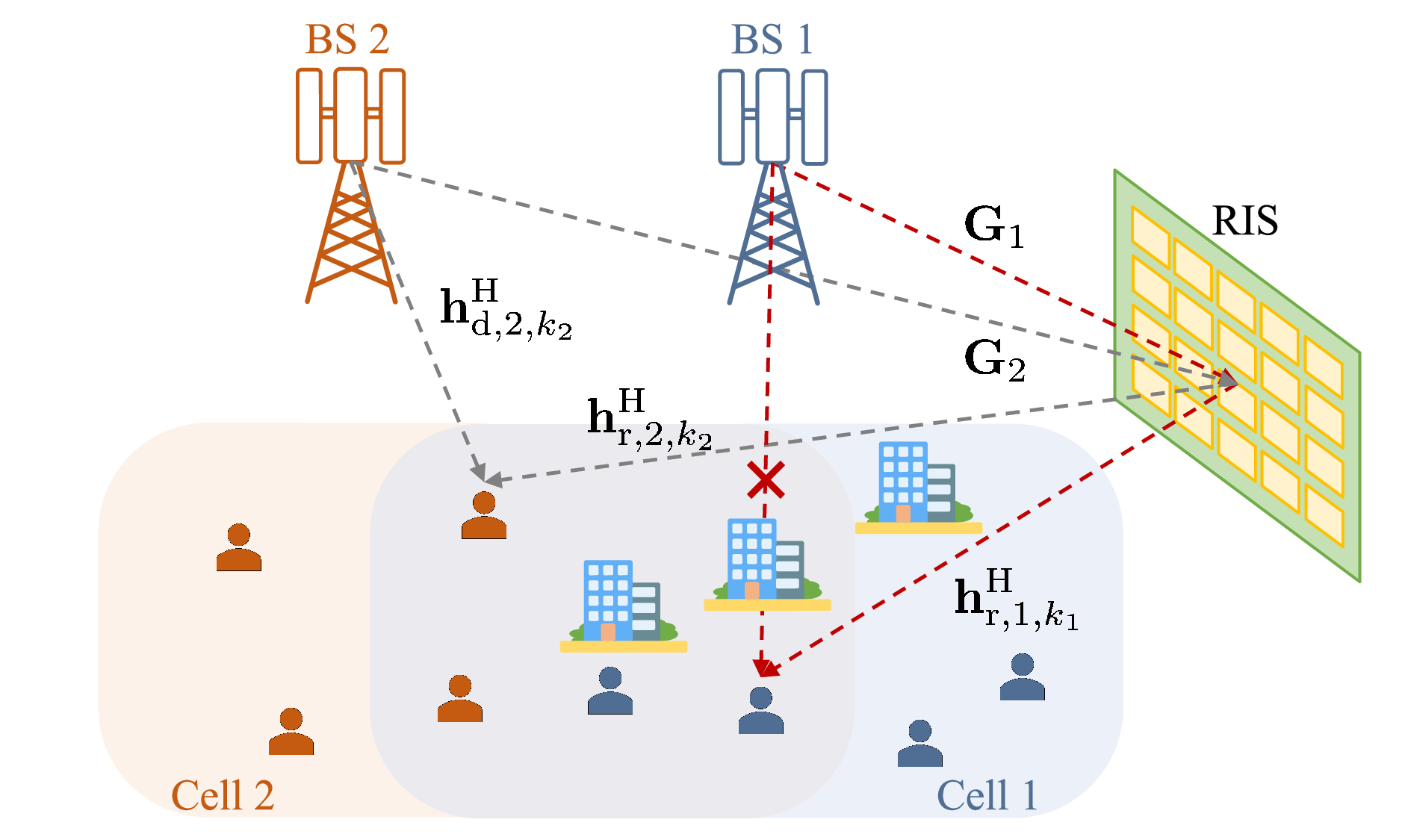}
	\caption{An example of realistic RIS-enabled cellular communication systems.}
	\label{system model figure}
\end{figure}

We consider a realistic cellular communication system as shown in Fig. \ref{system model figure} where two service providers operate one BS each for their serving cells,\footnote{To explicitly reveal the effect of deploying RIS for the scenario of interest, we assumed the simplified model featuring only two service providers and their serving cells. However, it is important to note that both the system model and the proposed RIS design can be easily extended to general cases involving more than two service providers as long as only one RIS is deployed in the~system.} and the BSs exploit different frequency ranges. For the $i$-th cell where $i\in\{1,2\}$, BS~$i$ deploys $N_i$ antennas to serve $K_i$ single-antenna users. We use the index $k_i$ to denote the $k$-th user in $\cK_i$ where $\cK_i \triangleq \{1_i,\cdots,K_i\}$ is the set of users served by BS $i$. We assume that the direct link channels between BS 1 and the users in $\cK_1$ are totally blocked by obstacles, and an RIS, connected via a control link to BS~1, with $M$ passive elements is deployed to overcome the blockage through the RIS~\cite{RIS_obstacle_1},~\cite{RIS_obstacle_2}.

In general, the communication system for the users in each cell will be separately designed by each service provider with different operating frequency ranges. However, the RIS reflects incoming signals regardless of frequency. This implies that the users in $\cK_2$ also experience \textit{uncontrolled} reflective signals through the RIS. As in Fig. \ref{system model figure}, the downlink channel between the $i$-th BS and the RIS is denoted by $\bG_i\in\mathbb{C}^{M \times N_i}$, and $\bh_{\mathrm{r},i,k_i}^\mathrm{H}\in\mathbb{C}^{1 \times M}$ implies the reflection channel between the RIS and the $k_i$-th user. The direct link channel between BS 2 and the $k_2$-th user is denoted by $\bh_{\mathrm{d},2,k_2}^\mathrm{H}\in\mathbb{C}^{1 \times N_2}$. Since the proposed design does not depend on a specific channel model, we do not assume any model here.

For the linear transmit beamforming at BS $i$, $\bs_i \triangleq \left[ s_{i,1_i},\cdots,s_{i,K_i}\right]^\mathrm{T}\in\mathbb{C}^{K_i \times 1}$ is the transmitted symbols following $\mathbb{E}\{\bs_i\bs_i^\mathrm{H}\}=\bI_{K_i}$, and $\bF_i \triangleq \left[\bff_{i,1_i},\cdots,\bff_{i,K_i} \right]\in\mathbb{C}^{N_i \times K_i}$ denotes the corresponding beamforming matrix satisfying the total power constraint $\mathrm{Tr}\left(\bF_i^\mathrm{H}\bF_i\right) \leq P_{\mathrm{T},i}$ with the maximum downlink transmit power $P_{\mathrm{T},i}$. Then, the downlink received signal at the $k_1$-th user in $\cK_1$ is given by
\begin{align}
	y_{1,k_1}=\left(\bh_{\mathrm{r},1,k_1}^\mathrm{H}\bPhi\bG_1\right)\sum_{k'_1\in \cK_1}\bff_{1,k'_1}s_{1,k'_1}+n_{1,k_1},
	\label{received signal cell 1}
\end{align}
where $n_{1,k_1}\sim \cC\cN(0,\sigma^2_{1,k_1})$ denotes the additive Gaussian noise. The reflection coefficient matrix at the RIS is defined by the $M \times M$ diagonal matrix $\bPhi \triangleq \diag(\bphi^\mathrm{H})$ with $\bphi=\left[\phi_1,\cdots,\phi_M\right]^\mathrm{T}$ where $\vert\phi_m\vert=1$ for $m=1,\cdots, M$. 

Considering practical hardware implementation of the RIS, some existing works discovered that even with the same RIS settings, the reflection coefficients will be different for the signals at different frequencies \cite{multi_cell_RIS_6_selective, RIS_frequency_selective_1, RIS_frequency_selective_2}. However, the proposed architecture in \cite{RIS_frequency_selective_1} suggests that the linear phase shift responses according to voltage with the same slope can be achieved when the gap between two frequency ranges is not too large, and the similar results are observed in \cite{RIS_frequency_selective_2}. In practice, taking the frequency spectrum of LTE as an example, the frequency gap between two neighboring service providers is about 20 MHz, which is small enough to assume a constant phase offset for the reflection coefficients of RIS to the reflective signals at different frequency ranges.\footnote{Although there may be amplitude degradation for each element, some advanced hardware techniques can compensate for it \cite{multi_cell_RIS_6_selective}. Therefore, we neglect the practical amplitude response and just assume the unit-norm constraint for each element in this paper.} Then, by denoting the reflection coefficient matrix for the reflective channels at cell~2 as $\bar{\bPhi} = e^{j\theta} \bPhi$ with a constant phase offset~$\theta$, the received signal at the $k_2$-th user in $\cK_2$ is 
\begin{align}
	y_{2,k_2}&=\left(\bh_{\mathrm{d},2,k_2}^\mathrm{H}+\bh_{\mathrm{r},2,k_2}^\mathrm{H}\bar{\bPhi}\bG_2\right)\sum_{k'_2\in \cK_2}\bff_{2,k'_2}s_{2,k'_2}+n_{2,k_2}, \notag \\
	&=\left(\bh_{\mathrm{d},2,k_2}^\mathrm{H}+e^{j \theta}\bh_{\mathrm{r},2,k_2}^\mathrm{H}{\bPhi}\bG_2\right)\sum_{k'_2\in \cK_2}\bff_{2,k'_2}s_{2,k'_2}+n_{2,k_2},
	\label{received signal cell 2}
\end{align}
where $n_{2,k_2}\sim \cC\cN(0,\sigma^2_{2,k_2})$ is the additive Gaussian noise. 

Following the received signal models, the achievable rate $R_{1,k_1}$ for the $k_1$-th user in $\cK_1$ is given by
\begin{align}
	R_{1,k_1} &= \log_2(1+\gamma_{1,k_1}), \notag \\
	\gamma_{1,k_1} &= \frac{\left\vert\left(\bh_{\mathrm{r},1,k_1}^\mathrm{H}\bPhi\bG_1\right)\bff_{1,k_1} \right\vert^2}{\sum_{k'_1 \neq k_1}^{K_1}\left\vert\left(\bh_{\mathrm{r},1,k_1}^\mathrm{H}\bPhi\bG_1\right)\bff_{1,k'_1} \right\vert^2+\sigma_{1,k_1}^2},
\end{align}
where $\gamma_{1,k_1}$ implies SINR at the $k_1$-th user, and the sum-rate of the users in $\cK_1$ is obtained by $R_1=\sum_{k_1\in \cK_1} R_{1,k_1}$. Similarly, the achievable rate $R_{2,k_2}$ for the $k_2$-th user in $\cK_2$ is
\begin{align}
	R_{2,k_2} &= \log_2(1+\gamma_{2,k_2}), \notag \\
	\gamma_{2,k_2} &= \frac{\left\vert\left(\bh_{\mathrm{d},2,k_2}^\mathrm{H}+e^{j \theta}\bh_{\mathrm{r},2,k_2}^\mathrm{H}{\bPhi}\bG_2\right)\bff_{2,k_2} \right\vert^2}{\sum_{k'_2 \neq k_2}^{K_2}\left\vert\left(\bh_{\mathrm{d},2,k_2}^\mathrm{H}+e^{j \theta}\bh_{\mathrm{r},2,k_2}^\mathrm{H}{\bPhi}\bG_2\right)\bff_{2,k'_2} \right\vert^2+\sigma_{2,k_2}^2},
\end{align}
where $\gamma_{2,k_2}$ is SINR at the $k_2$-th user. The sum-rate of the users in $\cK_2$ is obtained by $R_2=\sum_{k_2\in \cK_2} R_{2,k_2}$.

To investigate the scenario of interest for the first time, we assume that BS 1 can obtain perfect channel knowledge for all links to design $\bF_1$ and $\bPhi$ that will be elaborated in Section~\ref{section: proposed balancing reflection design}. In contrast, we consider the worst-case scenario for BS 2 that it only has the direct link channel information of the users in $\cK_2$, i.e., $\{\bh_{\mathrm{d},2,k_2}\}_{k_2 \in \cK_2}$, for constructing $\bF_2$. With this assumption, we can explicitly observe the effect of RIS on the users in $\cK_2$ when BS 2 does not have any capability to handle the uncontrolled channels through the RIS due to the lack of any knowledge related to the RIS.

\section{Proposed Balancing Reflection Design} \label{section: proposed balancing reflection design}

As can be clearly seen through the received signal model in~\eqref{received signal cell 2}, the RIS still affects the users in $\cK_2$. When the reflection coefficients of RIS are designed only considering the users in~$\cK_1$, the RIS can cause undesired performance degradation to the users in $\cK_2$ due to the limited channel knowledge at BS 2, i.e., only the direct link channels are known. To resolve this issue, in this section, we propose a balancing reflection design for the RIS by considering its effect on the channels of the users in $\cK_1$ and $\cK_2$ simultaneously.

\subsection{Problem Formulation}
To design the RIS operation for our scenario of interest, it is crucial to set a proper performance metric. Unfortunately, it is difficult to quantify the undesired performance degradation of the users in $\cK_2$ by the RIS in terms of standard performance metrics such as the sum-rate $R_2$. In addition, simply minimizing $R_2$ would be highly impractical. Therefore, we focus on the reflective channels and the uncontrolled channels through the RIS for the users in $\cK_1$ and $\cK_2$.

To reveal the operation of RIS more explicitly, we first reformulate the reflective channel through the RIS at the $k_1$-th user in \eqref{received signal cell 1} as 
\begin{align}
	\bh_{\mathrm{r},1,k_1}^\mathrm{H}\bPhi\bG_1 &= \bphi^\mathrm{H}\diag\left(\bh_{\mathrm{r},1,k_1}^\mathrm{H} \right)\bG_1 =\bphi^\mathrm{H}\bA_{1,k_1},
\end{align}
by defining $\bA_{1,k_1} \triangleq \diag\left(\bh_{\mathrm{r},1,k_1}^\mathrm{H} \right)\bG_1 \in \mathbb{C}^{M \times N_1}$. Then, for all users in $\cK_1$, the total gain of reflective channels is given~by 
\begin{align}
	\sum_{k_1 \in \cK_1}\bphi^\mathrm{H}\bA_{1,k_1} \bA_{1,k_1}^\mathrm{H}\bphi &= \bphi^\mathrm{H}\left(\sum_{k_1 \in \cK_1}\bA_{1,k_1} \bA_{1,k_1}^\mathrm{H}\right)\bphi, \notag \\ &= \bphi^\mathrm{H} \tilde{\bA}_1 \bphi,
\end{align}
where we define $\tilde{\bA}_1 \triangleq \sum_{k_1 \in \cK_1}\bA_{1,k_1} \bA_{1,k_1}^\mathrm{H}\in \mathbb{C}^{M \times M}$ as the total reflective channel through the RIS for the users in $\cK_1$. Similarly, the uncontrolled channel by the RIS at the $k_2$-th user in \eqref{received signal cell 2} can be reformulated as 
\begin{align}
	e^{j\theta}\bh_{\mathrm{r},2,k_2}^\mathrm{H}\bPhi\bG_2
	&=e^{j\theta}\bphi^\mathrm{H}\diag\left(\bh_{\mathrm{r},2,k_2}^\mathrm{H} \right)\bG_2=e^{j\theta}\bphi^\mathrm{H}\bA_{2,k_2},	
\end{align}
by defining $\bA_{2,k_2} \triangleq \diag\left(\bh_{\mathrm{r},2,k_2}^\mathrm{H} \right)\bG_2\in \mathbb{C}^{M \times N_2}$. The gain of this channel is given as
\begin{align}
	(e^{j\theta}\bphi^\mathrm{H}\bA_{2,k_2})(e^{j\theta}\bphi^\mathrm{H}\bA_{2,k_2})^\mathrm{H}&= e^{j\theta}\bphi^\mathrm{H}\bA_{2,k_2} \bA_{2,k_2}^\mathrm{H}\bphi e^{-j\theta}, \notag \\
	&=\bphi^\mathrm{H}\bA_{2,k_2} \bA_{2,k_2}^\mathrm{H}\bphi,
\end{align} 
where any effect of $\theta$ is disappeared. Then, the total gain of uncontrolled channels for the users in $\cK_2$ is  
\begin{align}
	\sum_{k_2 \in \cK_2}\bphi^\mathrm{H}\bA_{2,k_2} \bA_{2,k_2}^\mathrm{H}\bphi &= \bphi^\mathrm{H}\left(\sum_{k_2 \in \cK_2}\bA_{2,k_2} \bA_{2,k_2}^\mathrm{H}\right)\bphi, \notag \\ &= \bphi^\mathrm{H} \tilde{\bA}_2 \bphi,
\end{align}
where $\tilde{\bA}_2 \triangleq \sum_{k_2 \in \cK_2}\bA_{2,k_2} \bA_{2,k_2}^\mathrm{H}\in \mathbb{C}^{M \times M}$ implies the total uncontrolled channel by the RIS for the users in $\cK_2$. 

Since the original purpose of the RIS is to overcome the blockage of the users in $\cK_1$, maximizing $\bphi^\mathrm{H} \tilde{\bA}_1 \bphi$ will be one of the main design objectives. At the same time, for the users in $\cK_2$, mitigating the undesired performance degradation by the RIS is important, and the RIS also should be developed to minimize its effect on cell 2, i.e., minimize $\bphi^\mathrm{H} \tilde{\bA}_2 \bphi$. With this design philosophy, the two different objectives can be formulated as the problem $(\mathrm{P}1)$, which is given as 
\begin{align}
	(\mathrm{P}1): \quad &\max_{\bphi} \quad  \bphi^\mathrm{H} \left(\frac{\tilde{\bA}_1}{\Vert\tilde{\bA}_1\Vert_\mathrm{F}}-\lambda\frac{\tilde{\bA}_2}{\Vert\tilde{\bA}_2\Vert_\mathrm{F}} \right) \bphi \quad\quad \label{P1 objective function} \\
&\enspace \mathrm{s.t.} \quad \vert \bphi_m \vert=1, \enspace \forall m\in\{1,\cdots,M \}. \quad\quad \label{P1 unit modulus constraint}
\end{align}
In \eqref{P1 objective function}, the relative weight between the two design objectives is controlled via the balancing parameter $\lambda \in \mathbb{R}^+$, i.e., large $\lambda$ implies more emphasis on minimizing the total uncontrolled channel gain by the RIS, and each total channel $\tilde{\bA}_i$ is normalized to effectively control the weight only through $\lambda$. For notational simplicity, we define
\begin{align}
	\bR(\lambda) \triangleq \frac{\tilde{\bA}_1}{\Vert\tilde{\bA}_1\Vert_\mathrm{F}}-\lambda\frac{\tilde{\bA}_2}{\Vert\tilde{\bA}_2\Vert_\mathrm{F}},
\end{align}
which is an $M \times M$ Hermitian matrix. Then, the reformulated problem $(\mathrm{P}1')$ is given as
\begin{align}
	(\mathrm{P}1'): \quad &\min_{\bphi} \quad -\bphi^\mathrm{H} \bR(\lambda) \bphi \label{P1 prime objective function} \\ 
	&\enspace \mathrm{s.t.} \quad \vert \bphi_m \vert=1, \enspace \forall m\in\{1,\cdots,M \}. \quad\quad \label{P1 prime unit modulus constraint}
\end{align}
\subsection{Algorithm Development}

Although the objective function of $(\mathrm{P}1')$ is expressed as the simple quadratic formula with respect to $\bphi$, unfortunately, the problem is still non-convex due to the unit modulus constraints for the reflection coefficients as in \eqref{P1 prime unit modulus constraint}. To obtain an effective solution of $(\mathrm{P}1')$, we adopt the Riemannian manifold optimization method as in \cite{Riemannian_2, Riemannian_3} to effectively handle the non-convex constraints in~\eqref{P1 prime unit modulus constraint}. One can easily find that the element-wise unit modulus constraints in $\bphi$ form a complex circle manifold given as $\cM_{\mathrm{cc}}^M = \left\{ \bphi \in \mathbb{C}^{M \times 1} : \vert \phi_1 \vert = \cdots = \vert \phi_M \vert =1 \right\}$, which is a Riemannian manifold, and the objective function in \eqref{P1 prime objective function} is continuous and differentiable with respect to $\bphi$. Therefore, we can adopt the Riemannian conjugate gradient (RCG) algorithm, which operates with the following three key steps.


\subsubsection{Riemannian Gradient Computation} We denote the objective function in \eqref{P1 prime objective function} as $f(\bphi)=-\bphi^\mathrm{H} \bR(\lambda) \bphi$. At a point $\bphi^{(t)} \in \cM_{\mathrm{cc}}^M$ at the $t$-th iteration, the Riemannian gradient represents the direction of the greatest decrease of $f(\bphi)$ and is given by a tangent vector given by the orthogonal projection of the Euclidean gradient $\nabla f\left(\bphi^{(t)}\right)$. Then, the Riemannian gradient $\mathrm{grad} f\left(\bphi^{(t)}\right)$ is 
\begin{align}
	\mathrm{grad} f\left(\bphi^{(t)}\right) =\nabla f\left(\bphi^{(t)}\right) - \mathrm{Re} \left(\nabla f\left(\bphi^{(t)}\right) \odot \bphi^{(t)*} \right) \odot \bphi^{(t)}, \label{Riemannian gradient}
\end{align}
where the Euclidean gradient of the objective function in \eqref{P1 prime objective function} is simply obtained by $\nabla f(\bphi) = -\bR(\lambda) \bphi$. 

\subsubsection{Vector Transport} With $\mathrm{grad} f\left(\bphi^{(t)}\right)$ in hand, the search direction $\bd^{(t)}$ at $\bphi^{(t)}$ can be updated similarly with the CG method as in the Euclidean space. In general, $\bd^{(t+1)}$ and $\bd^{(t)}$ in manifold optimization would lie in the two different tangent spaces, and even simple operations such as summation of two vectors within different tangent spaces cannot be directly conducted. Still, the mapping between two tangent vectors, called transport, can overcome this problem. The vector transport for the manifold $\cM_{\mathrm{cc}}^M$ is defined as 
\begin{align}
	\cT_{\bphi^{(t)}\rightarrow \bphi^{(t+1)}}\left(\bd^{(t)}\right) \triangleq \bd^{(t)} - \mathrm{Re} \left(\bd^{(t)} \odot \bphi^{(t+1)*} \right) \odot \bphi^{(t+1)},
\end{align}
and the update rule for $\bd^{(t)}$ is given by 
\begin{align}
	\bd^{(t+1)} = -\mathrm{grad} f\left(\bphi^{(t+1)}\right) + \beta^{(t)}\cT_{\bphi^{(t)}\rightarrow \bphi^{(t+1)}}\left(\bd^{(t)}\right), \label{search direction}
\end{align}
where $\beta^{(t)}$ is chosen as the Polak-Ribiere parameter \cite{Riemannian_book}. 

\alglanguage{pseudocode}
\begin{algorithm}[t]
	\caption{Proposed balancing reflection coefficients design}
	\textbf{Initialization}
	\begin{algorithmic}[1]
		\State Initialize $\bphi^{(0)} \in \cM_{\mathrm{cc}}^M$, $\bd^{(0)}=-\mathrm{grad}f\left(\bphi^{(0)}\right)$, and $t=0$
	\end{algorithmic}
	\textbf{Iterative update}
	\begin{algorithmic}[1]
	\addtocounter{ALG@line}{+3}
		\Repeat
		\State Choose step size $\alpha^{(t)}$
		\State Find next point $\bphi^{(t+1)}$ by retraction in \eqref{retraction}
		\State Compute $\mathrm{grad} f\left(\bphi^{(t+1)}\right)$ according to \eqref{Riemannian gradient}
		\State Choose $\beta^{(t)}$
		\State Conduct vector transport to update $\bd^{(t+1)}$ as in \eqref{search direction}
		\State $t \leftarrow t+1$
		\Until Convergence
	\end{algorithmic}
	\textbf{Output:} $\tilde{\bphi}=\bphi^{(t)}$
\end{algorithm}

 \begin{figure}[t]
	\centering
	\includegraphics[width=0.9\columnwidth]{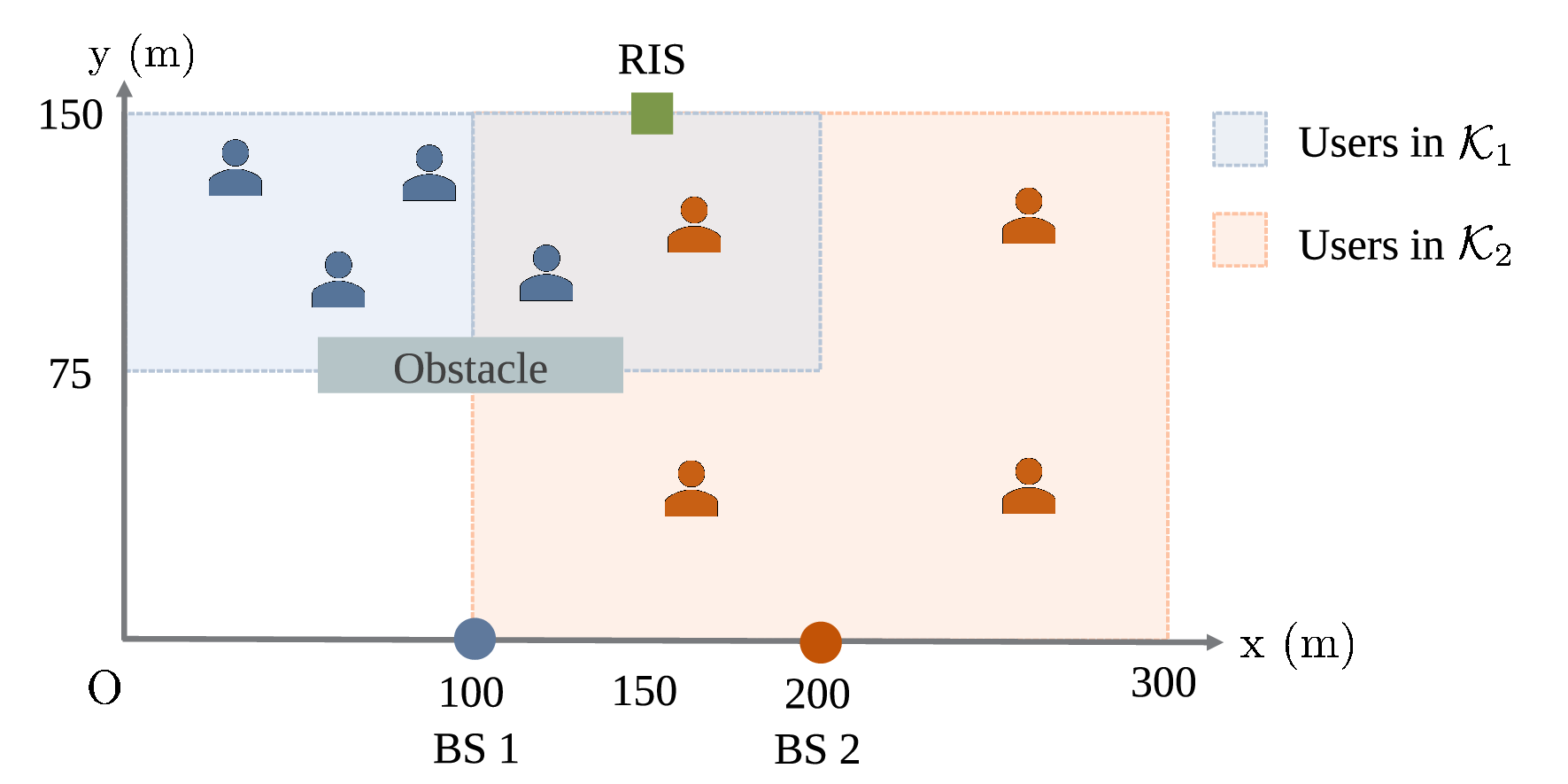}
	\caption{Positions of each BS and its serving area with RIS.}
	\label{location figure}
\end{figure}

\subsubsection{Retraction} After determining $\bd^{(t)}$, the retraction operation, which is a mapping from the tangent space onto the manifold, is applied to find the next destination on the manifold. With the Armijo backtracking line search step size $\alpha^{(t)}$ for $\bd^{(t)}$ \cite{Riemannian_book}, the updated point $\bphi^{(t+1)}$ is expressed as
\begin{align}
	\bphi^{(t+1)} = \cR\left(\bphi^{(t)} + \alpha^{(t)}\bd^{(t)} \right), \label{retraction}
\end{align}
where the retraction operation $\cR(\cdot)$ is defined by
\begin{align}
	\cR\left(\bphi\right) \triangleq \left[\frac{\phi_1}{\vert \phi_1 \vert},\cdots, \frac{\phi_M}{\vert \phi_M \vert}\right]^\mathrm{T}.
\end{align}

The overall process of the RCG algorithm is summarized in Algorithm 1 where the critical point convergence is guaranteed~\cite{Riemannian_book}. With the final effective solution $\tilde{\bphi}$, BS 1 designs $\bF_1$ based on the total channel information. For the computational complexity analysis of Algorithm 1, we first assume $N_1=N_2=N$ and $K_1=K_2=K$. The complexity of obtaining $\bR(\lambda)$ is given by $\cO(2 N M^2 K)$. Since the complexity of RCG algorithm is mainly dominated by computing the Euclidean gradient, which is given by $\cO(M^2)$, the overall required complexity of Algorithm 1 becomes $\cO(2 N M^2 K+IM^2 )$ where $I$ denotes the number of iterations~\cite{Riemannian_3}.

{\textit{Remark.} What we want to focus on in this work is the effect of deploying RIS for realistic cellular systems operated by different service providers, not the methodology for solving $(\mathrm{P}1')$. Although we adopted the Riemannian manifold optimization method in this work, the semi-definite relaxation method can be exploited as in \cite{RIS-SDP}, and a solution can be obtained by normalizing each element of the maximum eigenvector of~$\bR(\lambda)$.

\section{Numerical Results} \label{section: numerical results}

In this section, we evaluate the performance of proposed balancing reflection design at the RIS for realistic cellular communication systems. Considering a practical 20 MHz frequency gap, the constant phase offset is set as $\theta=\pi/6$  \cite{RIS_frequency_selective_2}. For each beamforming matrix $\bF_i$, we adopt the well-known signal-to-leakage-and-noise-ratio-based beamforming technique with the equal power allocation for simplicity\cite{SLNR}. Note that as we discussed in Section~\ref{section: system model}, $\bF_2$ is obtained only with the direct link channel information~$\{\bh_{\mathrm{d},2,k_2}\}_{k_2 \in \cK_2}$.

\begin{table}[t]
	\renewcommand{\arraystretch}{1.5} 
	\centering
	\caption{Rician fading channel parameters.}
	\begin{tabular}{| c | c | c | c |}
		\hline
		\multirow{2}*{Communication links} & \multirow{2}*{\shortstack{Path loss                                          \\exponents}} & \multirow{2}*{\shortstack{Rician \\ factors}} & \multirow{2}*{\shortstack{Number of\\NLoS paths}} \\ & & & \\
		\hline
		\multirow{1}*{BS-user direct}      & \multirow{1}*{$4.2$}                  & \multirow{1}*{$3 \text{ }\mathrm{dB}$} & \multirow{1}*{$8$} \\
		\hline
		\multirow{1}*{RIS-user reflection} & \multirow{1}*{$2.4$}                  & \multirow{1}*{$5 \text{ }\mathrm{dB}$} & \multirow{1}*{$4$} \\
		\hline
		\multirow{1}*{BS-RIS}              & \multirow{1}*{$2.5$}                  & \multirow{1}*{$5 \text{ }\mathrm{dB}$} & \multirow{1}*{$8$} \\
		\hline
	\end{tabular}
\end{table}

Considering a three-dimensional (3D) coordinate system, Fig. \ref{location figure} shows the positions of BSs, RIS, and users in the xy-plane where the users are uniformly distributed in each serving area. The height of BSs, RIS, and users are set as 15 m, 10 m, and 1 m, respectively. Assuming half-wavelength spacing, the antennas of BSs and elements of RIS are deployed in uniform planar array structures, which are aligned to the xz-plane. We consider $N_{i,\mathrm{ver}}$ vertical and $N_{i,\mathrm{hor}}$ horizontal antennas for BS $i\in\{1,2\}$ where $N_i = N_{i,\mathrm{ver}} \times N_{i,\mathrm{hor}}$. Similarly, $M_\mathrm{ver}$ vertical and $M_\mathrm{hor}$ horizontal elements are deployed for the RIS where $M=M_\mathrm{ver}\times M_\mathrm{hor}$. For all channel links, we adopt the practical Rician fading channel model, which contains one line-of-sight (LoS) path and multiple non-line-of-sight (NLoS) paths as in \cite{Rician_channel_model_1, Rician_channel_model_2}. The angles of LoS path are numerically obtained with the actual positions, and the angles of NLoS paths are randomly generated based on those of LoS path. The path-loss exponent, Rician factor, and number of NLoS paths for each link are given in Table I. The noise variance at each user is set as $\sigma_{1,k_1}^2=\sigma_{2,k_2}^2=-104$ dBm. To reveal the homogeneity~among the cells, we assume the same number of BS antennas and users, i.e., $N_1=N_2=N$ and $K_1=K_2=K$. Also, the maximum downlink transmit powers $P_{\mathrm{T},1}$ and $P_{\mathrm{T},2}$ are set to be equal to $P_\mathrm{T}$. We take the sum-rates $R_1$ and $R_2$ as the performance metric.

We compare the proposed design with the following baselines.
\begin{itemize}
	\item Conv. RIS: This implies the conventional RIS operating scenario such that the RIS is designed to maximize only the total reflective channel gain of the users in $\cK_1$ through the RIS by ignoring the users in $\cK_2$~\cite{RIS-SDP}. In this case, Algorithm 1 is applied with $\lambda=0$ for $(\mathrm{P}1')$.
	\item Rand. RIS: The phase shifts of all reflection coefficients are uniformly and randomly distributed in $[0, 2\pi )$.
	\item No-RIS: For cell 2, we consider this case that the users in $\cK_2$ only experience the direct channel links without any undesired performance degradation. It is possible to verify how the RIS deployed in cell 1 affects the performance of cell 2 by comparing it with this case.
\end{itemize}

Fig. \ref{sum_rate_vs_tx_power_N_44_M_816_K_4_lambda_20dB} shows the average sum-rate performance of the proposed design and baseline schemes according to the downlink transmit power $P_\mathrm{T}$. For cell 2, we can consider the No-RIS case as a virtual upper bound. Then, the performance gap between the No-RIS case and Conv. RIS case implies the undesired performance degradation when the RIS deployed in cell 1 is designed without any consideration of its effect on the other cell. In contrast, the proposed balancing RIS design takes the uncontrolled channels for the users in $\cK_2$ into account, and it shows higher sum-rate performance than the Conv. RIS and Rand. RIS cases. 

\begin{figure}[t]
	\centering
	\includegraphics[width=0.9\columnwidth]{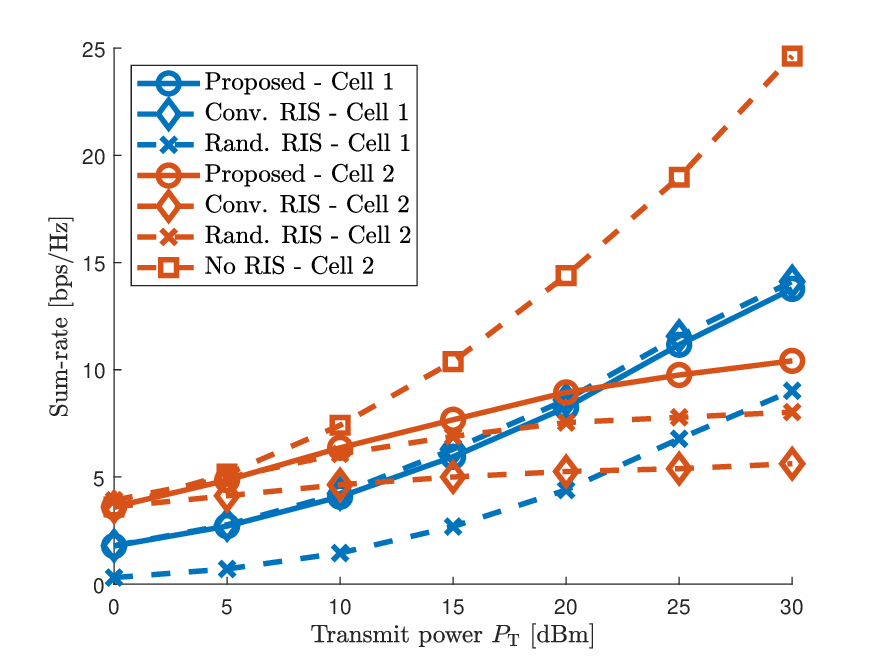}
	\caption{Average sum-rate performance according to $P_\mathrm{T}$ with $N=4\times 4$, $M=8 \times 16$, $K=4$, and $\lambda=20$ dB.}
	\label{sum_rate_vs_tx_power_N_44_M_816_K_4_lambda_20dB}	
\end{figure}

In Fig. \ref{sum_rate_vs_tx_power_N_44_M_816_K_4_lambda_20dB}, we can also observe that the sum-rate of proposed design is slightly lower than the Conv. RIS case for cell 1. This is because the RIS is not designed solely for the users in $\cK_1$.  Nevertheless, we want to emphasize that the proposed design is still effective since the performance improvement compared to the Conv. RIS case for cell~2 is much more prominent than the degradation for cell~1. This becomes much clearer in Fig.~\ref{sum_rate_vs_lambda_N_44_M_816_K_4_txp_30dBm} where the figure depicts the average sum-rate performance according to $\lambda$. The proposed design achieves dramatically increasing sum-rate performance for cell 2 as $\lambda$ increases, while the degradation for cell 1 is negligible until $\lambda=20$ dB. Furthermore, the results in Fig. \ref{sum_rate_vs_lambda_N_44_M_816_K_4_txp_30dBm} suggest that the proper value of $\lambda$ can be numerically optimized depending on the design purpose of overall communication system. For instance, if the system can compromise with some performance degradation for cell 1 and wants to minimize the undesired performance degradation for cell 2, a large $\lambda$ can be adopted and vice versa.   

\begin{figure}[t]
	\centering
	\includegraphics[width=0.9\columnwidth]{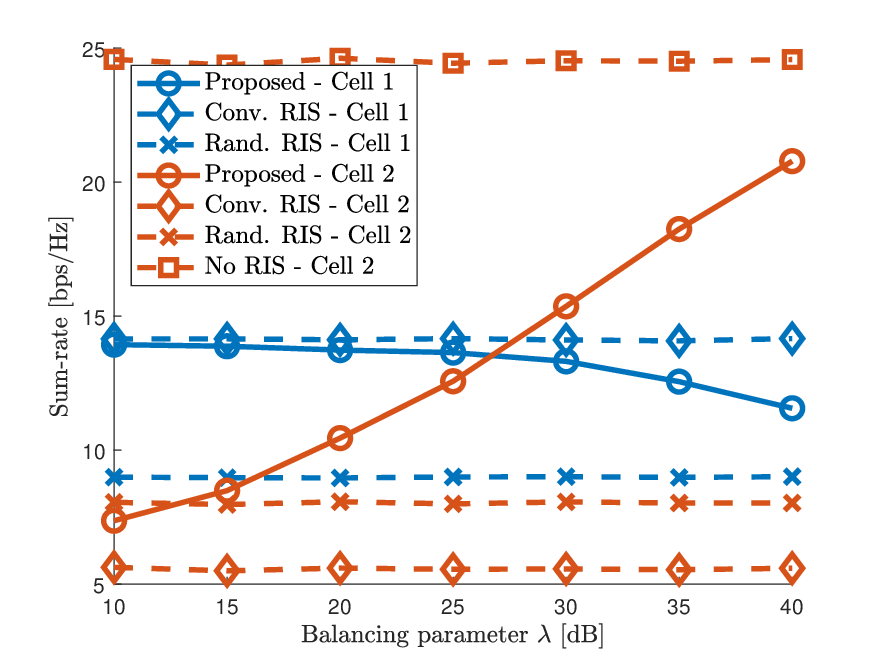}
	\caption{Average sum-rate performance according to $\lambda$ with $N=4\times 4$, $M=8 \times 16$, $K=4$, and $P_\mathrm{T}=30$ dBm.}
	\label{sum_rate_vs_lambda_N_44_M_816_K_4_txp_30dBm}
\end{figure}

\section{Conclusion} \label{section: conclusion}
We proposed the balancing RIS design for realistic cellular communication systems where each cell is operated by a distinct service provider using a different frequency range. By exploiting the Riemannian manifold optimization method, the reflection coefficients are carefully designed to maximize the total channel gain of the cell that deploys the RIS and to minimize the uncontrolled channel gain for the other cells simultaneously. Numerical results show that the proposed design with the proper balancing parameter can achieve high performance improvement for the cells that do not deploy the RIS with negligible degradation for the cell assisted by the~RIS. 

There are many interesting future research directions for our scenario of interest including the followings:
\begin{enumerate}
    \item \textbf{Joint design of beamforming and reflection coefficients}: Investigate the joint optimization of beamforming technique at the BSs and reflection coefficients at the RIS to improve the overall system performance.
    \item \textbf{Exploring more general system models}:
    \begin{itemize}
        \item Co-located BSs: Examine scenarios when BSs from different service providers are co-located, which could introduce new opportunities by exploiting high channel correlations.
        \item Multi-cell environments/multiple BSs for each service provider: Extend the analysis to systems with more than two cells or scenarios where each service provider operates multiple BSs. It could provide insights into network-wide performance and coordination strategies.
    \end{itemize}
    \item \textbf{Various channel knowledge assumptions}: Analyze scenarios with different channel knowledge assumptions. For instance, imperfect channel knowledge can be assumed at the BS that utilizes the RIS, and the other BSs may have partial channel information such as the second-order statistics of reflective channels through the RIS. This can reflect real-world environments more closely and provide a better understanding of practical limitations and strategies for improving system performance under channel uncertainty.
    \item \textbf{Extension to system deploying active RIS}: Investigate the effect of deploying active RIS, which can overcome the double path loss effect as in \cite{active_RIS}, on our scenario of interest.
\end{enumerate}

\bibliographystyle{IEEEtran}
\bibliography{refs_multi_cell_multi_band_RIS.bib}

\end{document}